\documentstyle[12pt,cmb,epsf]{article}

\begin{document}

\heading{The Doppler peaks from a generic defect}

\author{Jo\~ao Magueijo}
{DAMTP and MRAO, University of Cambridge, Cambridge, UK}{\quad}

\begin{abstract}{\baselineskip 0.4cm 
We investigate which of the exotic Doppler peak features 
found for textures and cosmic strings are generic novelties
pertaining to defects. We find that the ``out of phase'' 
texture signature is an accident. Generic defects, when they generate
a secondary peak structure similar to inflation, apply to it an additive
shift. It is not necessary for this shift to be ``out of phase''.
We also show which factors are responsible for the absence of secondary
oscillations found for cosmic strings. Within this general analysis
we finally consider the conditions under which topological defects and
inflation can be confused. It is argued that only $\Omega=1$ 
inflation and a defect with a horizon size coherence length
have a chance to be confused. Any other inflationary or defect model
always differ distinctly.\\}
\end{abstract}

\section{Introduction}
Recent work has addressed the hitherto virgin ground of Doppler peaks
induced by motivated topological defect scenarios. Textures have been
studied by \cite{neil} and \cite{ruth} with good qualitative agreement,
and cosmic strings were studied in \cite{us,us1,us2}. This work prompts
reflection on a more fundamental level. Inflationary perturbations have
always been open to the most general class of possibilities. For instance
one considered tilt, tensor modes, etc, long before  motivated inflationary
mechanisms were found to justify them. Defect theories on the contrary
have always been tied down to concrete models produced by specific 
patterns of spontaneous symmetry breaking. 
Here we shall depart from this (highly
positive) trend, and allow ourselves the same speculative freedom
inflationary theories enjoy. We shall therefore ask what
is a generic defect, and what are the generic Doppler peak features 
induced by a generic defect.

One metaphysical and one practical motivation assist this project. 
Firstly one would like to know how much of what has already been
found for textures and strings are generic defect novelties. For instance
it was found in \cite{neil,ruth} that  texture Doppler peaks appear out of 
phase with respect to the standard inflationary peaks. 
Is this a robust defect feature?
Or is it an accident pertaining to textures? 
In other words, could a generic defect apply an arbitrary shift
to  the Doppler peaks? This is an important question, as
topological defects and standard isocurvature perturbations (like the
ones discussed in \cite{hw}) are not at all the same thing, a fact often
overlooked. Therefore there is no reason why an out of phase signature
would have to be associated with defects. A second example is provided
by cosmic strings. Despite many uncertainties it has been
found that cosmic strings do not have secondary Doppler peaks.
This is a rather exotic feature, completely alien to inflationary
theories. Is this feature a robust prediction for a large class 
of defect models? And if so what are the controlling factors 
responsible for the opposite behaviour of strings and textures in this 
respect?

A second, perhaps more practical motivation for this type of work
lies in the conflict between inflation and defects. Are the inflationary
Doppler peaks proof of inflation, or could one in principle cook up
a defect which reproduced the inflationary Doppler peaks? Until this
question has been answered it is daylight robbery to claim that 
a measurement of, say, the CDM prediction for the $C_\ell$
spectrum, would prove inflation. 

A detailed and more serious discussion of this problem may be found
in \cite{us2}. Here we merely highlight the most entertaining aspects
of this work.

\section{The ontology of defects and inflation}\label{definfl}
We now focus on the basic assumptions of
inflationary and defect theories and  isolate  the most shocking
contrasting properties. 
We define the concepts of active and passive perturbations,
and of coherent and incoherent perturbations. In terms of 
these concepts inflationary perturbations are 
passive coherent perturbations.
Defect perturbations are active perturbations more or less
incoherent depending on the defect.

\subsection{Active and passive perturbations, and their different
perceptions of causality and scaling}
The way in which inflationary and defect perturbations come about
is radically different. Inflationary fluctuations 
were produced at a remote epoch, and were 
driven far outside the Hubble radius by  inflation.  The
evolution of these fluctuations is linear (until 
gravitational collapse becomes non-linear at late times), and we call
these fluctuations ``passive''.  Also, because
all scales observed today have been in causal contact since the onset
of inflation, causality does not strongly constrain the fluctuations
which result. In contrast, defect fluctuations are continuously seeded by
defect evolution, which is a non-linear process.
We therefore say these are ``active'' perturbations.  Also, the
constraints imposed by causality on defect formation  and evolution 
are much greater than  those placed on inflationary perturbations.

\subsubsection{Active and passive scaling}
The notion of scale invariance has different implications
in these two types of theory. For instance, a scale invariant gauge-invariant
potential $\Phi$ with dimensions $L^{3/2}$ has a power spectrum 
$$P(\Phi)=\langle |\Phi_{\bf k}|^2\rangle\propto k^{-3}$$ 
in passive theories (the Harrison-Zeldovich spectrum). 
This results from the fact that
the only variable available is $k$, and so the only spectrum one
can write down which has the right dimensions and does not have a 
scale is the Harrison-Zeldovich spectrum. 
The situation is different
for active theories, since time is now a variable.
The  most general counterpart to the Harrison-Zeldovich spectrum is  
\begin{equation}\label{scale}
P(\Phi) =  \eta^3F_{\Phi}(k\eta)
\end{equation}
where $F_{\Phi}$
is, to begin with, an arbitrary function of $x=k\eta$. All other
variables may be written as a product of a power of $\eta$, ensuring
the right dimensions, and an arbitrary function of $x$. 
Inspecting all equations it can be checked that it is possible to do
this consistently for all variables. All equations respect scaling
in the active sense.

\subsubsection{Causality constraints on active perturbations}\label{causal}
Moreover, active perturbations are constrained by causality, in the form of 
integral constraints \cite{trasch12,james}. These consist of 
energy and momentum conservation laws for fluctuations
in an expanding Universe. The integral constraints can be used to 
find the low $k$ behaviour of the perturbations' power spectrum, 
assuming their causal generation and evolution
\cite{traschk4}. Typically it is found that the causal creation
and evolution of defects requires that their energy $\rho^s$ and scalar
velocity $v^s$
be white noise at low $k$, but that the total energy fluctuations'
power spectrum is required to go like $k^4$. To reconcile these two facts
one is forced to consider the compensation. 
This is an underdensity in the matter-radiation  energy density
with a white noise low $k$ tail, correlated with the defect network
so as to cancel the defects' white-noise tail. When one combines the
defects energy with the compensation density, one finds that the 
gravitational potentials they generate also have to be
white noise at large scales \cite{us2}. Typically the scaling function
$ F_{\Phi}(k\eta)$ will start as a constant and decay as a power law
for $x=k\eta>x_c$. The value $x_c$ is a sort of coherence wavenumber
of the defect. The larger it is the smaller the defect is. For instance
$x_c\approx 12$ for cosmic strings (thin, tiny objects), whereas
$x_c\approx 5.5$ for textures (round, fat, big things). Sophisticated
work on causality \cite{james} has shed light on how small $x_c$
may be before violating causality. The limiting lower bound $x_c
\approx 2.7$  has been suggested.

Although we will not here have a chance to dwell on technicalities,
it should be stated that the rather general discussion presented above
is enough to determine the general form of the potentials for active 
perturbations. This has been here encoded in the single parameter $x_c$.
We shall see that $x_c$ will determine the Doppler peak position
for active perturbations. Doppler peaks are driven by the gravitational
potential, so it should not be surprising that the defect length scale
propagates into its potential, and from that  into the Doppler peaks'
position.

\subsection{Coherent and incoherent perturbations}
Active perturbations may also differ from
inflation in the way ``chance'' comes into the theory. 
Randomness occurs in inflation only when the initial 
conditions are set up. Time evolution is linear and
deterministic, and may be found by
evolving all variables from an
initial value equal to the square root of 
their initial variances. By squaring the
result one obtains the variables' variances at any time.
Formally this results from unequal time correlators of the form
\begin{equation}\label{2cori}
{\langle\Phi({\bf k},\eta)\Phi({\bf k'},\eta ')\rangle}=
\delta({\bf k}-{\bf k'})\sigma({\Phi}(k,\eta))\sigma
({\Phi}(k,\eta')),
\end{equation}
where $\sigma$ denotes the square root of the power spectrum $P$.
In defect models however, randomness may intervene in the time
evolution as well  as the initial conditions. 
Although deterministic in principle, 
the defect network evolves as a result of a 
complicated non-linear process.
If there is strong non-linearity, a given mode will be ``driven'' 
by interactions with the other modes in a way which will force
all different-time correlators to zero on a time scale
characterized by the ``coherence time'' $\theta_c(k,\eta)$.
Physically this means that one has to perform a new ``random'' draw 
after each coherence time in order to
construct a defect history \cite{us}. 
The counterpart to (\ref{2cori}) for incoherent perturbations is
\begin{equation}\label{pr0}
{\langle\Phi({\bf k},\eta)\Phi({\bf k'},\eta ')\rangle}=
\delta({\bf k}-{\bf k'}) P({\Phi}(k,\eta),\eta'-\eta)\; .
\end{equation}
For $|\eta'-\eta| \equiv |\Delta\eta|> \theta_c(k,\eta)$
we have $P({\Phi}(k,\eta),\Delta\eta)=0$. For $\Delta\eta=0$,
we recover the power spectrum $P({\Phi}(k,\eta),0)=P({\Phi}(k,\eta))$.

We shall label as coherent and incoherent
(\ref{2cori}) and (\ref{pr0})  respectively. 
This feature does not affect the Doppler peaks' position
but it does affect the structure of secondary oscillations.
An incoherent potential will drive the CMB oscillator incoherently,
and therefore it may happen that the secondary oscillations get
washed out as a result of incoherence.

\section{Generic defect Doppler peaks}
\begin{figure}[t]
\begin{center}
    \leavevmode
        {\hbox %
{\epsfxsize = 11cm\epsfysize=11cm
    \epsffile {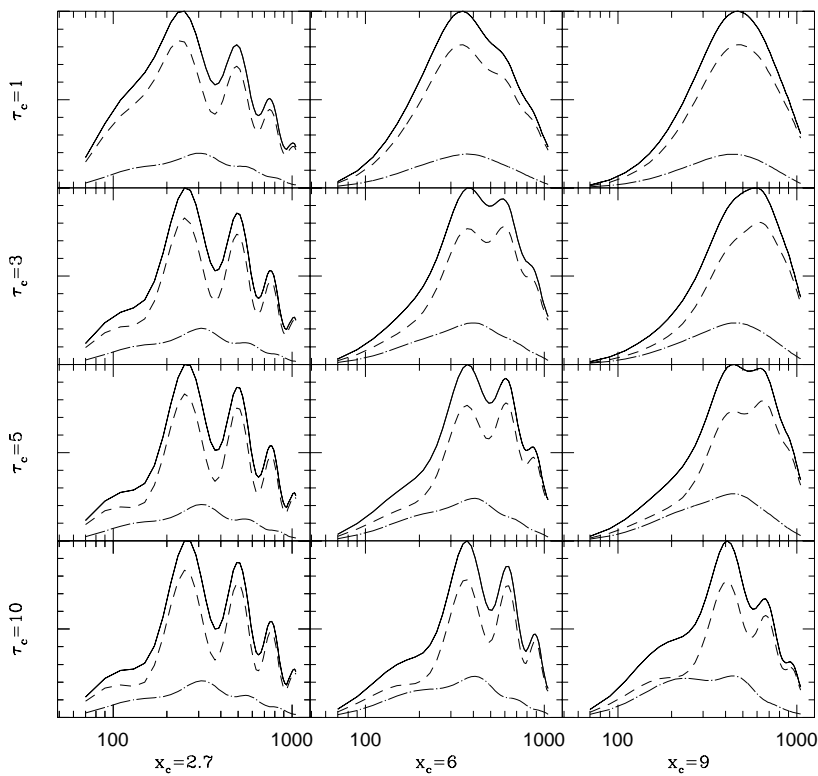} }}
\end{center}
\caption{$C_l$ spectra for a grid of models with various values
of $x_c$ (related to the defect coherence length) and 
$\theta_c\approx 2.35\tau_c$ (the defect coherence time). We have included
the monopole term (dash) and dipole term (point-dash), Silk damping,
and free-streaming. The monopole term is always dominant.}
\label{fig1}
\end{figure}

In Fig.~\ref{fig1} we 
show a grid of $C_\ell$ spectra functions of the two parameters
introduced above: $x_c$ (related to the defect coherence length) and 
$\theta_c$ (the defect coherence time). For the exact form of the stress
energy of these defects we refer the reader to \cite{us2}.

\subsection{The peaks position}
In general there may or may not be a system of secondary Doppler 
peaks. However if they exist, then their position is determined purely
by $x_c$.
For $x_c\approx 2.7$ (not impossible,
but probably unrealistic because it is 
very close to the smallest
turnover point allowed by causality \cite{james,causal-ngt}) 
the peaks are at the adiabatic
positions. As $x_c$ increases from the adiabatic position the peaks 
are shifted to smaller scales.
For $x_c\approx 5.4$ they are out of phase with the adiabatic peaks
(as in \cite{neil,ruth}). For $x_c>8.5$ the peaks start only
in the adiabatic secondary peaks region. For standard 
values of $\Omega_b$ and $h$ these three cases would place the
main ``Doppler peak'' at $l\approx 230$, $350$, and $500$, respectively. 
Therefore the placing of the peaks is {\it not} a generic
feature of active fluctuations. 
Active perturbations simply add an extra parameter on which the
Doppler peaks position is strongly dependent. In general we should expect
that for the same $\Omega$, $\Omega_b$, and $h$, active perturbations
will apply to the predicted CDM adiabatic peak position
a shift of the form
\begin{equation}
  \label{shift}
  l\rightarrow
l+{\eta_0\over \eta^*}{\left(x_c -{\pi{\sqrt3}\over 2}\right)}
\end{equation}
where $\eta_0$ and $\eta^*$ are the conformal times nowadays and at
recombination.
The secondary peaks' separation is not changed, in a first approximation.
This is to be contrasted with non-flat inflationary models 
where $C_l(\Omega=1)$ is taken into $C_{l\Omega^{-1/2}}$.
The defect shift is additive whereas the low-$\Omega$ shift 
is multiplicative, a striking difference that should always allow us to
distinguish between low $\Omega$ CDM and $\Omega=1$
high-$x_c$ defects. 

\subsection{Intensity of secondary oscillations}
The strength of the secondary
oscillations depends on both $x_c$ and $\theta_c$. For $x_c\approx
2.7$ there are secondary oscillations regardless of the exact
$\theta_c$ value. This is a confusing defect, as not only does it place
the Doppler peaks on the adiabatic position, but also the peak
structure is quite insensitive to the defect incoherence. 
For larger $x_c$ the secondary Doppler peaks survive only if 
the defect coherence time is much larger than $x_c$. This condition
seems unphysical for large $x_c$ so we expect realistic
defects with large $x_c$
not to have secondary oscillations.

This can be understood heuristically. Incoherence tends to erase secondary
oscillations. However each mode is active only for a period of scaling
time of the order of $x_c$. If $\theta_c>x_c$ then each mode is coherent
for longer than it is active, and so the defect is 
effectively coherent and the secondary oscillations are preserved. 
If on the contrary $\theta_c<x_c$ then the defect has time
to display its incoherence.  Large defects (small $x_c$) are not active
for long enough to display whatever reasonable incoherent properties
they may have. Very small defects (large $x_c$) 
are active for long enough for their
incoherence to be manifest, whatever reasonable coherence time they may have.

To help the reader to connect this general discussion with concrete
defect theories, here is a rough guide to the topography of Fig.~\ref{fig1}. 
Current understanding places the cosmic string models on 
the top right corner of Figure~\ref{fig1} (large $x_c$, $\tau_c$
smaller than 3). They should have a single peak well after the main
adiabatic peak. Textures fall
somewhere in the middle of the figure ($x_c$ around 6, coherence
time not yet measured). Their main peak should be out of phase
with the adiabatic peaks. This is an accident related to the $x_c$
value for textures, and not a robust defect feature. Texture  secondary
oscillations should exist but be softer than predicted by the
coherent approximation (used in \cite{neil,ruth}). How 
much softer depends on the exact value of the texture's $\theta_c$.
If their coherence time is of the same order as strings ($\tau_c\approx
3$) their secondary oscillation will be very soft.

\section{Confusing defects and inflation}
Given this state of affairs what are the chances of confusing
inflation and defect theories? The answer to this question depends
on $\Omega$, on the inflation side, and on $x_c$, on the defect side.
We have shown how only $\Omega=1$ inflation and $x_c\approx 2.7$
(the causal lower bound) have a chance to be confused. Any defect
with a larger $x_c$ is bound to cause great disarray in what has
come to be expected from Doppler peaks by inflationary trends.
Two novelties stand out. First, if defects preserve a structure of
secondary peaks, then this tends to be obtained from the inflationary
one by an additive shift in $l$, rather than a multiplicative shift
(as it happens for low $\Omega$ inflation). Second, defects may erase
the secondary peaks.
 
It remains as an open question the case $\Omega=1$ inflation vs 
$x_c=2.7$ defects. Our work, and also \cite{causal-ngt} suggests
that indeed in this case one may confuse defects and inflation.
The work in \cite{hw} seems to draw the opposite conclusion.
One would hope that this issue is clarified in the not too distant
future.

\acknowledgements{I would like to thank Andy Albrecht, Pedro Ferreira
and David Coulson for the very enjoyable collaboration leading up to this
work. I am indebted to Kim Baskerville for reading this manuscript,
and to Anne Davis for partial financial support.
I should finally thank Ruth Durrer for ensuring my mental sanity
during these Rencontres.}

\vfill
\end{document}